\documentclass[
]{ceurart}

\usepackage{listings}
\usepackage{graphicx}
\usepackage{url}
\usepackage[htt]{hyphenat}
\usepackage{tabularx}
\usepackage{xcolor}
\usepackage{caption}
\usepackage{tcolorbox}
\usepackage{float}
\tcbuselibrary{listings,skins,breakable}

\definecolor{myblue}{RGB}{137, 171, 209}
\definecolor{myred}{RGB}{237, 120, 122}
\definecolor{mygreen}{RGB}{89, 176, 92}
\definecolor{mygray}{RGB}{204, 204, 204}

\lstdefinelanguage{SPARQL}{
  morekeywords={SELECT,WHERE,FILTER,EXISTS,NOT,OPTIONAL,UNION,GRAPH,FROM,ORDER,BY,GROUP,HAVING,ASK,CONSTRUCT,DESCRIBE,PREFIX,a},
  sensitive=true,
  morecomment=[l]{\#},
  morestring=[b]"
}

\lstdefinestyle{sparql}{
  language=SPARQL,
  basicstyle=\ttfamily\footnotesize,
  keywordstyle=\color{blue!70!black}\bfseries,
  stringstyle=\color{green!40!black},
  commentstyle=\color{gray},
  breaklines=true,
  columns=fullflexible,
  showstringspaces=false,
  frame=single,
  framerule=0.3pt,
  xleftmargin=0pt,
  xrightmargin=0pt,
  numbers=left,
  numberstyle=\tiny\color{gray},
  stepnumber=1,
  numbersep=5pt
}

\begin{document}

\copyrightyear{2025}
\copyrightclause{Copyright for this paper by its authors.
  Use permitted under Creative Commons License Attribution 4.0
  International (CC BY 4.0).}

\conference{CLAIRVOYANTS Workshop, Co-located with JURIX 2025}

\title{DAOnt: A Formal Ontology for EU Data Act Compliance}

\author[1]{Sheyla Leyva-Sánchez}[%
orcid=0009-0007-9762-4045,
email=sheyla.leyva.sanchez@upm.es,
]
\cormark[1]
\author[1]{Fabian Linde}[%
orcid=0009-0004-0856-8036,
email=fabian.linde@upm.es,
]
\author[1]{Meem Arafat Manab}[%
orcid=0000-0002-2336-4160,
email=meem.manab@upm.es,
]
\author[1]{María Poveda-Villalón}[%
orcid=0000-0000-0000-0000,
email=m.poveda@upm.es,
]
\author[1]{Víctor Rodríguez-Doncel}[%
orcid=0000-0003-1076-2511,
email=vrodriguez@fi.upm.es,
]

\address[1]{Universidad Politécnica de Madrid, Madrid, Spain}
\cortext[1]{Corresponding author.}

\begin{abstract}
The EU Data Act establishes comprehensive rules governing data access and sharing across business-to-consumer (B2C), business-to-business (B2B), and business-to-government (B2G) contexts. This paper presents a comprehensive ontology for the EU Data Act, enabling reasoning over data sharing agreements through machine-readable representations. The DAOnt ontology reuses elements from three established ontologies, LKIF-Core, ODRL, and DPV, to capture the normative structure of the Data Act. 

The ontology captures the main concepts and relationships in the Regulation, and it also operationalises three articles to facilitate compliance checking: Article 4(1) (B2C user access rights), Article 8(6) (B2B trade secret exceptions) and Article 19(2)(a) (B2G competitive use prohibitions).

The ontology supports compliance checking through SPARQL queries that return obligations, permissions, and prohibitions, allowing organisations to verify whether data-sharing agreements meet the requirements of the EU Data Act and to assess conditions such as FRAND obligations. By representing key legal concepts in RDF, our work helps bridge the gap between the legal provisions of the Data Act and their computational interpretation. The complete ontology, along with example instances and queries, is available online.
\end{abstract}

\begin{keywords}
  EU Data Act \sep
  Legal Ontology \sep
  Compliance Verification 
\end{keywords}

\maketitle

\section{Introduction}

The EU Data Act\footnote{Regulation (EU) 2023/2854 of the European Parliament and of the Council of 13 December 2023 on harmonised rules on fair access to and use of data (Data Act).} has become applicable on September 12, 2025, introducing a comprehensive regulatory framework for data access and sharing across the European Union. The regulation addresses three primary scenarios: business-to-consumer (B2C) data sharing where users gain access to data generated by their connected products (Chapter II, Articles 3-7), business-to-business (B2B) data sharing requiring user authorisation and fair, reasonable, and non-discriminatory (FRAND) terms (Chapter III, Articles 8-13), and business-to-government (B2G) data sharing under exceptional need circumstances (Chapter V, Articles 14-22).

Organisations face significant compliance challenges due to the complexity of the Regulation. The Data Act contains over 50 articles with intricate interdependencies: B2B sharing requires user authorisation per Article 8(4), must satisfy FRAND conditions per Articles 8-12, yet allows trade secret exceptions per Article 8(6). The manual compliance check is error-prone and does not scale, particularly for small and medium enterprises (SMEs) that lack dedicated legal resources. Data space architectures such as GAIA-X require interoperable contracts, while cloud providers need automated policy enforcement mechanisms.

The gap between legal text and computational enforcement creates practical barriers to compliance. Natural language provisions require human interpretation, leading to inconsistent implementations between organisations and jurisdictions. Without machine-readable representations, compliance checking remains a manual, costly process vulnerable to human error. Furthermore, existing compliance tools focus predominantly on GDPR and AI Act requirements, leaving the Data Act, despite its September 2025 applicability date, largely unaddressed by automated compliance solutions.

Machine-readable regulatory frameworks offer several advantages: automated reasoning can determine which obligations apply to specific scenarios, semantic queries can identify contract violations, and formal verification can validate policy compliance before deployment. Such frameworks enable affordable compliance tools for SMEs and support the interoperability requirements of European data spaces.

This paper addresses the research question: \textit{Can compliance checking with the EU Data Act requirements be automated through formal ontology and semantic reasoning?}

Our contribution is threefold. First, we present a comprehensive ontology for the EU Data Act, which integrates three established standards (LKIF-Core, ODRL, and DPV) to formally represent regulatory requirements. Second, we demonstrate proof-of-concept formalizations for Articles 4(1), 8(6), and 19(2)(a), showing how different deontic modalities (mandatory obligations in B2C, permissive exceptions in B2B, and absolute prohibitions in B2G) can be captured and reasoned over. Third, we provide executable reasoning examples using SPARQL queries over RDF-formatted knowledge bases, demonstrating practical automated compliance checking capabilities.

\subsection*{Paper Structure}
The remainder of the paper is structured as follows. Section~2 reviews related work on legal ontologies, data governance and data sharing ontologies, compliance approaches, and regulatory compliance ontologies, identifying the research gap that motivates our contribution. Section~3 presents the methodology, including the ontology engineering process and the integration of external vocabularies. Section~4 introduces the proof of concept, detailing the formalisation of Articles~4(1), 8(6), and 19(2)(a) and demonstrating automated reasoning examples in B2C, B2B and B2G compliance scenarios. Section~5 evaluates the ontology in terms of coverage, expressiveness, and interoperability. Finally, Section~6 concludes the article and Section~7 outlines the directions for future work.

\section{Related Work}
The development of an ontology for the EU Data Act builds upon two complementary research lines: the use of ontologies to represent legal and regulatory knowledge, and the formalisation of compliance mechanisms enabling automated or semi-automated assessment of normative requirements. The first line concerns the design of conceptual models that capture legal concepts such as rights, obligations, and permissions, providing a shared semantic foundation for interoperability and reasoning. These models are typically expressed using Semantic Web standards such as RDF and OWL, which ensure compatibility with existing ecosystems and support the representation of machine-readable normative structures. The second focusses on computational approaches to verify whether data processing or sharing activities comply with legal norms, using technologies such as SPARQL, SHACL, logic programming, or rule-based systems. This section reviews previous work in both areas, highlighting their relevance to the modelling of data-sharing obligations and justifying the design choices adopted in our ontology. 
 
\subsection{Legal ontologies}
Early legal ontologies were designed primarily as conceptual frameworks to represent the structure of legal knowledge rather than as operational data models. Their main purpose was to facilitate understanding, interoperability, and knowledge sharing within the legal domain, rather than to support good metadata or even machine-executable compliance checking. This focus on conceptual modelling distinguished them from later ontologies, such as those used in data governance and compliance, which aim to be directly operationalised through technologies like OWL reasoners, SHACL, or SPARQL.

Comprehensive surveys by Casellas \cite{casellas} and Rodrigues et al. \cite{rodrigues} provide an overview of this evolution. Casellas' foundational work in 2011 describes the emergence of legal ontologies as efforts to formalise legal concepts, case structures, and argumentation patterns, often within the contexts of knowledge management and document retrieval. In contrast, the more recent survey emphasises the shift toward lightweight, interoperable ontologies integrated with Semantic Web standards and designed to support automation, such as compliance checking and legal analytics.

Among the first and most influential initiatives, LKIF-Core (the Core Ontology of the Legal Knowledge Interchange Format)  \cite{lkif} stands out as a foundational ontology to represent legal reasoning and normative structures. Developed within the ESTRELLA project, LKIF-Core provides an upper-level schema for capturing concepts such as norms, acts, and roles, inspired by legal theory and deontic logic. Although not originally intended as an operational compliance model, LKIF-Core established a reusable conceptual foundation that has informed subsequent vocabularies. Its modular and extensible design continues to make it a reference point for aligning new legal ontologies with established conceptual categories.

LegalRuleML \cite{legalruleml} complements ontology-based representations with a formal markup language to express legal norms, rules, and reasoning structures. While LKIF-Core provides a conceptual vocabulary, LegalRuleML offers a syntactic and logical framework to encode the prescriptive aspects of law--obligations, permissions, prohibitions, and exceptions--using an XML-based serialisation aligned with RuleML \cite{ruleml}. Its primary objective is to enable the interchange of machine-readable legal rules across systems, preserving their temporal and defeasible characteristics. However, unlike lightweight ontologies expressed in RDF, LegalRuleML was not conceived as a data model for operational use within Linked Data environments, but rather as a rule representation language supporting formal reasoning. For this reason, it bridges conceptual and computational layers, yet its adoption has remained limited outside specialised rule-based reasoning systems. 

\subsection{Ontologies for Data Governance and Data Sharing}
The Open Digital Rights Language (ODRL) is a W3C specification, provided as two Recommendations \cite{odrl1}\cite{odrl2}, that provides a flexible and extensible policy expression language to represent permissions, prohibitions, and obligations associated with the use of digital assets. Originally conceived to manage usage rights in digital content distribution, ODRL has evolved into a general-purpose vocabulary to express policies that govern access, sharing, and usage of data and services. The activity in the domain of data spaces, closely related to our data act, is actually burgeoning.
ODRL adopts an RDF-based model, enabling interoperability within the Semantic Web ecosystem and allowing policies to be linked to resources, parties, and actions. Although not specifically designed for legal compliance, the normative structure of ODRL -- centred on the concepts of permission, prohibition, and duty -- makes it particularly suitable for modelling data-sharing agreements and other legally relevant policies and has been widely reused and extended in data governance and privacy-related vocabularies.

DaPReCo (Data Protection Regulatory Compliance Ontology) \cite{dapreco} is a formal ontology designed to represent the legal concepts and obligations of the General Data Protection Regulation (GDPR), with a focus on enabling rule-based compliance assessment . Based on LegalRuleML and FOL-based formalizations, DaPReCo encodes the logical structure of GDPR provisions, allowing automated reasoning over obligations, rights, and legal bases for data processing. Beyond DaPReCo, several other ontologies have sought to capture the semantics of the GDPR at varying levels of abstraction and operationality. 

The Data Privacy Vocabulary\footnote{\url{https://w3id.org/dpv/}} (DPV), developed under the W3C Data Privacy Vocabularies and Controls Community Group, provides a lightweight RDF-based model for representing data processing purposes, legal bases and data subject rights, emphasising interoperability and practical deployment. Other initiatives-such as GConsent \cite{gconsent} or PrOnto \cite{pronto} focus on narrower aspects such as consent modelling or personal data categories. Collectively, these ontologies illustrate a clear evolution from conceptual representations of privacy norms toward operational models that support machine-readable compliance verification.

\subsection{Compliance approaches}
Several conceptual approaches have been proposed to formalise and operationalise legal and regulatory compliance. \href{https://dalicc.net}{DALICC} (Data Access and Licence Interoperability for Content and Contracts) extends ODRL to model licencing and data-sharing policies in a way that supports interoperability across domains, bridging normative concepts with actionable constraints. In contrast, ontologies such as DaPReCo demonstrate how compliance checks can be performed by encoding GDPR rules into formal representations that allow reasoning engines to infer whether a given scenario satisfies obligations and permissions. Another widely used strategy involves SHACL (Shapes Constraint Language), which validates RDF data against predefined constraints, providing a declarative and relatively lightweight mechanism for ensuring compliance with structural or policy requirements. Some approaches go further by transforming ontology-based representations into other formal logic systems, such as first-order logic, Prolog, or Answer Set Programming (ASP), enabling advanced automated reasoning and defeasible rule evaluation. Finally, SPARQL offers a simpler yet effective means to query RDF representations directly, allowing organisations to verify specific obligations, prohibitions, or permissions without the overhead of full logic-based reasoning systems. Each of these approaches balances expressivity, computational complexity, and ease of integration, shaping the design choices for operational compliance ontologies.

\subsection{Regulatory Compliance Ontologies}

Automated compliance check approaches span multiple paradigms. Semantic logic-based approaches include First-Order Logic representations and frameworks such as I-SNACC that achieve 95.2\% precision and 100\% recall \cite{wu2023invariant}. SHACL-based compliance checking shows increasing adoption for validating regulatory requirements. Policy language–based approaches leverage ODRL and LegalRuleML to formalise rights and obligations in a machine-interpretable way. An ODRL Compliance Profile has also been proposed to demonstrate effective coverage when combined with DPV \cite{de2019odrl}. Recent innovations include the use of Defeasible Logic Programming to handle contradictions and incompleteness, Answer Set Programming with event calculus formalisms, and hybrid approaches combining machine learning and rule-based reasoning.

AI Act ontologies emerged rapidly following the regulation's adoption, including AIRO (AI Risk Ontology, 2022\cite{golpayegani2022airo}) providing OWL2 representation of AI system risks, VAIR (Vocabulary of AI Risks, 2023) for risk assessments, and TAIR (Trustworthy AI Requirements Ontology, 2024) modelling AI Act clauses and ISO standards.

\subsection{Research Gap: The Missing Data Act Ontology}

Despite the Data Act entering into force on January 11, 2024, with core provisions applying on September 12, 2025, no dedicated Data Act ontology currently exists. While DPV v2.0 includes Data Act support on its roadmap, this work remains incomplete. This represents an urgent gap for organisations that must comply without machine-readable specifications or automated compliance checking tools.

Current ontologies cannot express critical Data Act scenarios: connected product data rights, multi-tier data sharing chains (user → data holder → third party), conditional compensation mechanisms, FRAND term verification, switching rights for cloud services, or exceptional need provisions for public sector access. Our work addresses this gap through the first comprehensive formalisation of the Data Act.

\section{Methodology}

Traditional compliance verification relies on manual legal review or ad-hoc procedural code, both of which scale poorly as regulatory obligations become more numerous, interconnected, and dynamic. Ontologies provide a more robust alternative by formalising legal provisions into a machine-interpretable representation that reduces ambiguity and supports consistent, repeatable reasoning.

In our approach, the first stage consists of a comprehensive reading of the Data Act and internal discussions on how to formalise its key concepts, obligations, and legal actors. Based on this analysis, the regulation's requirements are modelled as explicit semantic structures with well-defined legal concepts and their normative relations. This semantic formalisation captures the regulatory logic in a form that software systems can reliably process. SPARQL queries then operationalise these structures, identifying compliance-relevant patterns with minimal procedural code.

This methodology transforms compliance assessment from a subjective, manual task into an automated and auditable process, improving transparency, scalability, and maintainability. As regulations evolve, ontology-based representations also offer a flexible foundation that facilitates continuous monitoring without extensive re-engineering.

\subsection{Ontology Engineering Process}

The development process follows the NeOn methodology~\cite{suarez-figueroa2012neon}, an iterative scenario-based framework well suited to modelling complex legal domains. We conducted a systematic requirement elicitation process through three complementary activities. 
First, a close reading and detailed annotation of the EU Data Act (Regulation 2023/2854) identified the main actors, core concepts, and their relationships throughout the regulation. 
Second, we formulate competency questions and organise them into two categories: general CQs ensuring core conceptual coverage (e.g. ``Who shares data with whom, and under which agreement?''), and article-specific CQs supporting the proof-of-concept formalisation (e.g., ``Which data holders violate Article~4(1) by failing to provide the user-requested data from connected products?''). The complete catalogue is provided in Table~\ref{tab:competency-questions}. The full Ontology Requirements Specification Document (ORSD) is included in Annex~\ref{annex:orsd}. Third, we performed an interoperability analysis to maximise reuse of established vocabularies—importing foundational legal concepts from LKIF-Core and extending data governance notions using DPV and ODRL—thereby avoiding redundant or conflicting definitions.

Tool-supported refinement was carried out throughout the iterative modelling cycle. Chowlk~\cite{chavez-feria2022chowlk} was used to automatically transform conceptual diagrams into formal OWL representations, and selected patterns from the Ontology Design Patterns catalogue ensured reusable and well-structured modelling decisions. Validation and refinement were performed through a consistency check with OWL reasoners (HermiT and ELK), complemented by SPARQL queries that verified that all competency questions were answerable. 

\begin{figure}[h]
    \centering
    \includegraphics[width=1\linewidth]{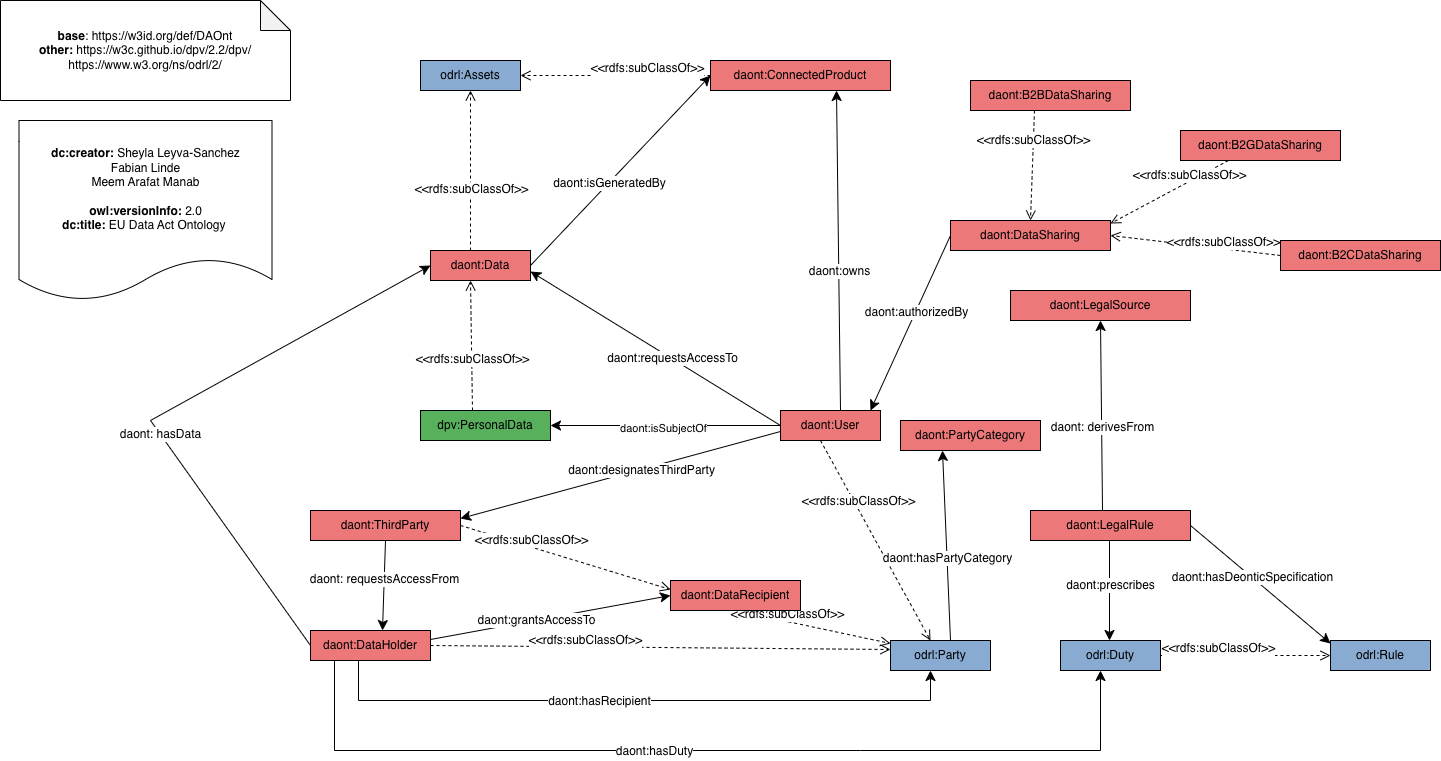}
    \caption{A top-level diagram of the ontology, depicting most relevant classes. (\textcolor{myred}{Red} tiles signify classes from the \textit{DAOnt} ontology, \textcolor{myblue}{blue} from \textit{ODRL}, \textcolor{mygreen}{green} from \textit{DPV})}
    \label{fig:placeholder}
\end{figure}

For documentation, we used Widoco~\cite{garijo2017widoco}, which generates human-readable specifications, metadata, and publication-ready HTML documentation. Finally, the ontology was published following the best practices of W3C using OnToology~\cite{chaves2013ontoology}, guaranteeing persistent hosting, versioning, and automated quality evaluation.\footnote{Public URI: \url{https://w3id.org/def/daont}}%
\footnote{Development repository: \url{https://github.com/oeg-upm/DAOnt}}

\begin{table}[h]
\centering
\renewcommand{\arraystretch}{1.2}
\begin{tabular}{|m{6cm}|m{6cm}|}
\hline
\cellcolor[HTML]{EFEFEF}\textbf{CQG1. General CQs (9 CQs)} 
& 
\cellcolor[HTML]{EFEFEF}\textbf{CQG2. CQs Related to PoC Articles (3 CQs)} 
\\ \hline

CQ1. Who shares data with whom, and under what agreement? \newline
CQ2. What actions has a party performed? \newline
CQ3. What obligations does a party have? \newline
CQ4. Who manufactured a product? \newline
CQ5. Who uses a product? \newline
CQ6. What service does a product provide? \newline
CQ7. When does a legal rule apply? \newline
CQ8. Where does data come from? \newline
CQ9. Who holds the data?
&

CQ10. Which data holders have violated Article 4(1) by failing to provide requested data from connected products? \newline
CQ11. Which data holders have violated Article 8(6) by refusing data sharing without providing trade secret justification? \newline
CQ12. Which public sector bodies have violated Article 19(2)(a) by developing competing products or services using data obtained through B2G data sharing?
\\ \hline

\end{tabular}
\caption{Competency questions grouped by general coverage (CQG1) and article-specific requirements for the proof of concept (CQG2).}
\label{tab:competency-questions}
\end{table}

\section{Proof of Concept: Three Regulatory Articles}

In this section, we present a proof of concept that illustrates how DAOnt enables automated compliance checking across different regulatory contexts of the Data Act. We analyse three representative articles (4(1){\footnote{Data holders shall make readily available data, as well as the relevant metadata necessary to interpret and use those data, accessible to the user without undue delay, of the same quality as is available to the data holder, easily, securely, free of charge, in a comprehensive, structured, commonly used and machine-readable format and, where relevant and technically feasible, continuously and in real-time}, 8(6) \footnote{An obligation to make data available to a data recipient shall not oblige the disclosure of trade secrets.}, 19(2)(a) \footnote{A public sector body, the Commission, the European Central Bank, a Union body or a third party receiving data under this Chapter shall not use the data or insights about the economic situation, assets and production or operation methods of the data holder to develop or enhance a connected product or related service that competes with the connected product or related service of the data holder} ), each corresponding to a distinct compliance modality—mandatory obligations, conditional permissions, and absolute prohibitions—and spanning the B2C, B2B, and B2G settings of the regulation. These examples demonstrate how the ontology captures normative structures with sufficient expressiveness to support automated reasoning. Table~\ref{tab:three-articles} provides a comparative overview of the formal representations used in each case.

\newcolumntype{Y}{>{\raggedright\arraybackslash}X}

\begin{table}[htbp]
\centering
\caption{Comparison of three Data Act articles}
\label{tab:three-articles}
\small
\begin{tabularx}{\textwidth}{|p{2.2cm}|Y|Y|Y|}
\hline
\textbf{Aspect} & \textbf{Article 4(1) B2C} & \textbf{Article 8(6) B2B} & \textbf{Article 19(2)(a) B2G} \\
\hline
\textbf{Legal Text} & "shall make data available without undue delay" & "not required to make data available where data constitute trade secrets" & "shall not use data to develop competing products" \\
\hline
\textbf{Deontic Type} & Obligation (\path{odrl:Duty}) & Permission (\path{odrl:Permission}) & Prohibition (\path{odrl:Prohibition}) \\
\hline
\textbf{Condition} & User requests data access & B2B request + trade secret & None (unconditional) \\
\hline
\textbf{Action} & \emph{provide\-Data\-Action} & \emph{refuse\-Data\-Access\-Action} & \emph{Use\-Data\-To\-Develop\-Competing\-Product} \\
\hline
\textbf{Exception} & None & \emph{contains\-Trade\-Secret\-Condition} & None \\
\hline
\textbf{Assignee} & Data holder (must act) & Data holder (may refuse) & Public sector body (cannot act) \\
\hline
\end{tabularx}
\end{table}

\subsection{Automated Reasoning Examples Across B2C, B2B, and B2G}

To demonstrate compliance verification capabilities, we developed six contract examples instantiating both compliant and non-compliant scenarios across the three data sharing contexts: B2C, B2B, and B2G. The class instances representing these six scenarios are available in the GitHub repository\footnote{\url{https://github.com/oeg-upm/DAOnt/tree/main/compliance-checks/contracts}}.

\vspace{5mm}

\textbf{B2C Context (Code Snippet ~\ref{lst:b2c-missing-obligation}).} 
In the violation scenario, \texttt{charlie} (a \texttt{ConsumerUser}) owns \texttt{smartWatch1}, which generates \texttt{charlieHealthData}. Charlie requests access to these data, but \texttt{watchManufacturer} (a \texttt{DataHolder} and \texttt{Manufacturer}) fails to perform any \texttt{DataProvision} action, thereby violating the mandatory obligation in Article~4(1). Data sharing is governed by \texttt{contract\_charlie}.  

To detect this violation automatically, the SPARQL query in Listing~\ref{lst:b2c-missing-obligation} searches for B2C data-sharing cases where a \texttt{ConsumerUser} requests access to data linked to a product they own or use, and checks whether the associated \texttt{DataHolder} performs the required \texttt{DataProvision} action. The \texttt{FILTER NOT EXISTS} block encodes the notion of a \emph{missing obligation}: if no \texttt{DataProvision} action is found, the pattern is flagged as non-compliant. The corresponding compliant scenario satisfies the obligation, and therefore the query does not return violations.

\lstset{style=sparql}
\begin{lstlisting}[caption={B2C: Missing Obligation}, label={lst:b2c-missing-obligation}, float=htbp]
SELECT ?holder ?data 
WHERE {
  ?sharing a da:B2CDataSharing;
           da:governedBy ?c .
  ?c dpv:hasRecipient ?user .
  ?user a da:ConsumerUser;
        da:ownsOrUses ?product;
        da:requestsAccessTo ?data.
  ?holder a da:DataHolder;
          dpv:hasData ?data .
  FILTER NOT EXISTS {
    ?holder da:performsLegalAction ?provision .
    ?provision a da:DataProvision .
  }
}
\end{lstlisting}

\begin{figure}[h]
    \centering
    \includegraphics[width=0.9\linewidth]{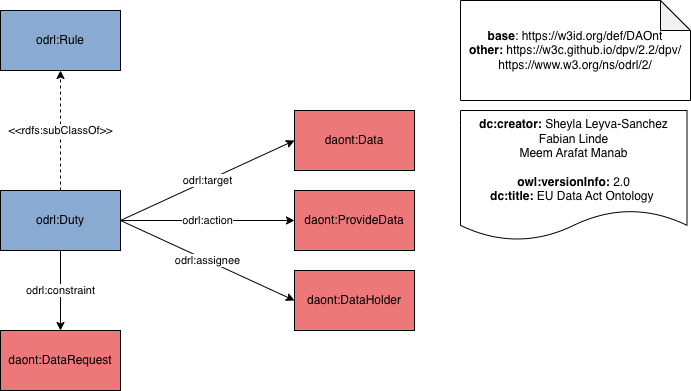}
    \caption{Diagram depicting the principal classes and properties involved in DA Art. 4(1), based on the ODRL design pattern. (\textcolor{myred}{Red} tiles signify classes from the \textit{DAOnt} ontology, \textcolor{myblue}{blue} from \textit{ODRL})}
    \label{fig:placeholder}
\end{figure}

\textbf{B2B Context (Code Snippet ~\ref{lst:b2b-refusal}).} 
In the violation scenario, \texttt{factoryOwnerAcme} (an \texttt{EnterpriseUser}) owns \texttt{industrialRobot1}, which generates \texttt{robotData1}. The factory owner authorises \texttt{autoRepair} (an \texttt{AftermarketServiceProvider} and \texttt{DataRecipient}) to access these data through \texttt{agreement247}, governed by \texttt{contract247}, which specifies FRAND-compliant terms (\texttt{frand247} with \texttt{isFair}, \texttt{isReasonable}, and \texttt{isNonDiscriminatory} all set to true). Despite these conditions, \texttt{robotManufacturer} (a \texttt{DataHolder}) refuses to share the data without providing a trade secret justification, violating the conditional sharing requirement in Article~8(6). The compliant scenario provides access under the same FRAND conditions and includes a valid justification whenever disclosure is refused.

To detect this violation automatically, the SPARQL query in Listing~\ref{lst:b2b-refusal} searches for B2B data-sharing cases where a \texttt{DataHolder} does not perform the required \texttt{DataProvision} action and simultaneously fails to provide a \texttt{containsTradeSecret} justification. The first \texttt{FILTER NOT EXISTS} block encodes the absence of the mandatory action, while the second captures the absence of the legally permitted exception. When both conditions hold, the query precisely identifies the non-compliant behaviour described in Article~8(6).

\begin{lstlisting}[caption={B2B: Refusal Without Trade Secret Justification}, label={lst:b2b-refusal}, float=htbp]
SELECT ?holder ?recipient 
WHERE {
  ?sharing a da:B2BDataSharing;
           da:governedBy ?c;
           da:authorizedBy ?user .
  ?user da:ownsOrUses ?product .
  ?product da:generatesData ?data .
  ?holder a da:DataHolder;
          dpv:hasData ?data .
  ?c dpv:hasRecipient ?recipient .
  FILTER NOT EXISTS {
    ?holder da:performsLegalAction ?provision .
    ?provision a da:DataProvision .
  }
  FILTER NOT EXISTS {
    ?data da:containsTradeSecret ?s .
  }
}
\end{lstlisting}

\begin{figure}[h]
    \centering
    \includegraphics[width=0.9\linewidth]{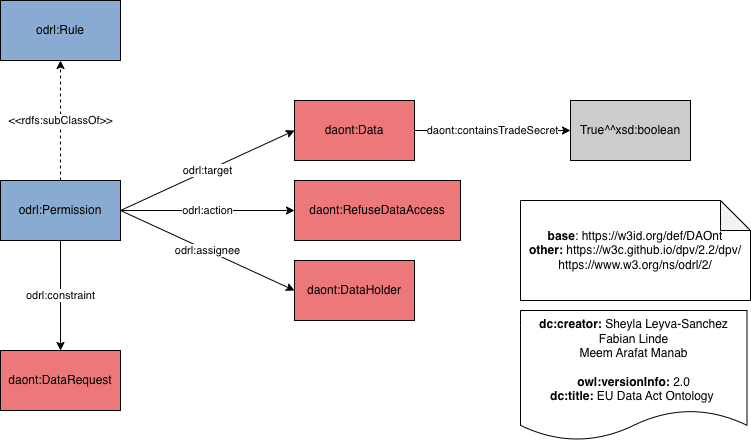}
    \caption{Diagram depicting the principal classes and properties involved in DA Art. 8(6), based on the ODRL design pattern. (\textcolor{myred}{Red} tiles signify classes from the \textit{DAOnt} ontology, \textcolor{myblue}{blue} from \textit{ODRL}, \textcolor{mygray}{gray} signifies XML Schema Definitions)}
    \label{fig:placeholder}
\end{figure}

\textbf{B2G Context (Code Snippet ~\ref{lst:b2g-prohibited-action}).}
In the violation scenario, \texttt{gonzalo} (a \texttt{ConsumerUser}) owns \texttt{healthMonitor1}, which generates \texttt{gonzaloHealthData}. The \texttt{healthAuthority} (a \texttt{PublicSectorBody} and \texttt{DataRecipient}) legitimately requests access to these data from the \texttt{healthDeviceManufacturer} under \texttt{publicHealthEmergency2024}, an instance of \texttt{ExceptionalNeed} authorising B2G sharing for pandemic monitoring. This exchange is governed by \texttt{contract191}. However, after obtaining the data, the authority performs \texttt{competitiveProductDevelopment1} (a \texttt{UseDataToDevelopCompetingProduct} action), thus using the data to develop a competing health application—an explicit violation of the absolute prohibition in Article~19(2)(a). The compliant scenario demonstrates lawful use restricted strictly to the declared public interest purpose.

To detect this violation automatically, the SPARQL query in Listing~\ref{lst:b2g-prohibited-action} searches for B2G data-sharing cases where a \texttt{PublicSectorBody} performs an action classified as \texttt{UseDataToDevelopCompetingProduct}. Unlike the B2C and B2B contexts, which require detecting \emph{missing obligations} or \emph{missing exceptions}, Article~19(2)(a) defines an absolute prohibition. For this reason, the query simply checks for the presence of the prohibited action; any match directly corresponds to non-compliance according to the regulation.

\begin{lstlisting}[caption={B2G: Prohibited Action}, label={lst:b2g-prohibited-action}, float=htbp]
SELECT ?publicBody ?action 
WHERE {
  ?sharing a da:B2GDataSharing;
           da:governedBy ?c .
  ?c dpv:hasRecipient ?publicBody .
  ?publicBody a da:PublicSectorBody .
  ?holder a da:DataHolder;
          dpv:hasData ?data;
          dpv:hasRecipient ?publicBody .
  ?publicBody da:performsAction ?action .
  ?action a da:UseDataToDevelopCompetingProduct .
}
\end{lstlisting}

\begin{figure}[ht]
    \centering
    \includegraphics[width=0.9\linewidth]{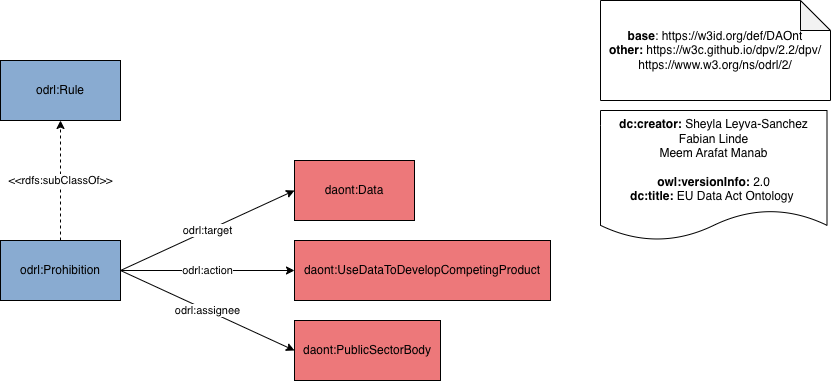}
    \caption{Diagram depicting the principal classes and properties involved in DA Art. 19(2)(a), based on the ODRL design pattern. (\textcolor{myred}{Red} tiles signify classes from the \textit{DAOnt} ontology, \textcolor{myblue}{blue} from \textit{ODRL})}
    \label{fig:placeholder}
\end{figure}

These queries illustrate how SPARQL operationalises the semantic structures defined in DAOnt to support automated compliance checking. By encoding mandatory obligations, conditional permissions, and absolute prohibitions as graph patterns, queries leverage the formal semantics of the ontology to detect violations directly from the data. This shows the practical value of combining ontology-driven modelling with query-based reasoning: compliance verification becomes scalable, transparent, and auditable, and regulatory requirements can be enforced without bespoke procedural logic.

\subsection{Automated Compliance Verification Engine}

These compliance verification queries are designed to detect Data Act violations through automated reasoning over RDF knowledge graphs. Such compliance checking may be needed in various contexts where data sharing occurs under regulatory obligations—for example, in data space infrastructures like Gaia-X\footnote{In Gaia-X, these are ``Compliance Services'' within the Trust Framework.}, sectoral platforms for health or mobility, or enterprise data sharing APIs. In these environments, a compliance module monitors activities by creating RDF instances for relevant events (contracts, data requests, data provisions) representing actual cases with real parties, datasets, and timestamps. Instance creation is hybrid: systems automatically capture events from APIs and smart contracts, while humans or LLMs judge concepts that cannot be formalised symbolically—like whether a delay is ``undue'' (Article 4) or if products ``compete'' (Article 19(2)(a)). SPARQL queries are then executed periodically or events are triggered to detect violations. Critically, this relies on the Closed World Assumption (CWA): \texttt{FILTER NOT EXISTS} interprets missing instances as violations, treating absence as non-compliance—contrasting with OWL's Open World Assumption where missing information means unknown. Since Data Act violations occur through inaction, the CWA reasoning is essential for compliance auditing.

To demonstrate this compliance verification approach, we developed a first implementation prototype in Python, using the RDFLib library to manage and query the knowledge graph. This approach enables the efficient loading of the foundational DAOnt schema along with the synthetic contract instances into a unified RDF graph. By executing the SPARQL queries for compliance (as detailed in the preceding section) directly against this graph, we were able to automatically detect instances of non-compliance. To validate the system's efficiency, we implemented a prototype compliance dashboard using Streamlit. This tool transforms the output of semantic reasoning into user-friendly, auditable reports that demonstrate several key advantages as follows: 

\begin{itemize}
    \item Efficiency: The application processes all six synthetic contracts in milliseconds, confirming the high throughput of the ontology-driven approach. This high processing speed contrasts sharply with the time and manual effort a human auditor would require to review contracts against regulatory text.

    \item Interpretable Results: Results are inherently transparent due to their foundation in SPARQL graph pattern matching on the DAOnt. The system provides a direct, traceable explanation by identifying the specific SPARQL rule and the exact entities involved in the violation.

    \item Auditable Oversight: The prototype facilitates immediate, interactive feedback for regulatory oversight by linking factual contractual data directly to the formal legal requirements encoded in DAOnt.
    
\end{itemize}

The operational prototype can be accessed at \href{https://daont-verify.streamlit.app/}{https://daont-verify.streamlit.app/}

\section{Evaluation and Discussion}

\subsection{Coverage Assessment}

We evaluated ontology coverage against the Data Act's core chapters using OnToology's automated documentation pipeline. The current implementation covers 18 articles in three primary contexts: B2C user access rights (Chapter II, Articles 3-7), B2B data sharing with FRAND conditions (Chapter III, Articles 8-13), and B2G exceptional need provisions (Chapter V, Articles 14-22). Articles 4(1), 8(6), and 19(2)(a) serve as representative formalizations demonstrating the modalities of obligation, permission, and prohibition, respectively. The remaining gaps include cloud switching mechanisms (Chapter VI, Articles 23-31), interoperability requirements (Chapter VII, Articles 32-34), and smart contract provisions (Chapter VIII, Article 30).

\subsection{Expressiveness Analysis}

OOPS! (OntOlogy Pitfall Scanner) detected no critical errors in the ontology structure. Minor pitfalls include missing domain/range declarations for 7 properties (P11 warning) and absence of inverse relationships for 4 object properties (P13 warning), both intentional design decisions to maintain compatibility with DPV's lightweight RDFS approach rather than full OWL2-DL constraints. The ontology successfully expresses complex deontic scenarios including conditional permissions (trade secret exceptions requiring \texttt{containsTradeSecretCondition}), multi-party chains (user → data holder → third party recipient), temporal constraints (Art. 4(1) "without undue delay"), and compensation mechanisms (Art. 12 FRAND terms).

\subsection{Interoperability Benefits, limitations and challenges}

Integration with four established standards enables multiple deployment scenarios. DPV alignment allows direct reuse in existing GDPR compliance systems (e.g., consent management platforms already using DPV vocabularies). ODRL compatibility supports data space implementations like GAIA-X requiring machine-readable policies. SPARQL queries execute on standard triple stores (GraphDB, Apache Jena, Virtuoso) without custom reasoning engines. LKIF-Core alignment facilitates future integration with legal reasoning systems that model Hohfeldian relationships. OnToology-generated documentation (HTML, diagrams, metadata) reduces adoption barriers through comprehensive reference materials.


Three primary limitations constrain current applicability. First, semantic ambiguity in legal text requires human interpretation for edge cases—Article 8(6) "trade secrets" lacks precise boundaries, requiring domain experts to instantiate \texttt{containsTradeSecret} predicates. Second, dynamic policy evolution remains unaddressed; contract modifications require manual knowledge base updates rather than automated policy versioning. Third, enforcement mechanisms lie outside ontology scope—while SPARQL queries detect violations, actual enforcement (notifications, sanctions, dispute resolution) requires external system integration. Validation with legal practitioners and industry stakeholders remains ongoing, with preliminary feedback indicating usability challenges for non-technical users requiring simplified tooling layers.

\section{Conclusions}

This work introduces the first formal ontology designed specifically to support compliance with the EU Data Act, addressing a critical gap in the landscape of machine-readable regulatory frameworks. By reusing and aligning concepts from ODRL, DPV, and LKIF-Core, DAOnt provides a coherent semantic model to represent rights, obligations, permissions, and prohibitions in data sharing agreements in B2C, B2B, and B2G contexts.

The proof-of-concept scenarios demonstrate that semantic modelling combined with SPARQL reasoning enables practical, fine-grained compliance checking. Executable queries successfully detect missing obligations, unjustified refusals, and prohibited actions, illustrating how legal requirements can be operationalised directly on top of ontology-driven representations. This confirms the viability of ontologies as a foundation for transparent, auditable, and scalable compliance verification.

Beyond immediate use cases, DAOnt lays the groundwork for future tools that support affordable compliance automation, interoperable contractual templates for data spaces, and policy-aware data governance infrastructures. As organisations prepare for the enforcement of the Data Act, ontology-based approaches such as the one presented here offer a promising pathway toward robust, trustworthy, and machine-processable regulatory compliance.

\section{Future Work}

Several directions remain open to expand the ontology and its compliance verification capabilities. 
First, we plan to conduct structured consultations with legal experts to refine the modelling of rights, obligations, permissions, and exceptions, ensuring that the ontology captures the nuances of legal interpretation and domain-specific practice. 
Second, the current proof-of-concept focusses on three representative articles; future work will expand the compliance checker to cover a broader range of provisions from the Data Act, including cross-cutting obligations, sector-specific rules, and complex exceptional-need scenarios. 
Third, we aim to explore the semi-automatic generation of Data Act–compliant contractual templates driven by our ontology, enabling machine-generated contracts to serve as executable test cases and reusable reference examples for automated compliance verification.

\begin{acknowledgments}
This research is funded by the European Union's Horizon 2020 programme under the Marie Skłodowska-Curie grant agreement No. \href{https://cordis.europa.eu/project/id/101169409}{101169409} (HARNESS) and project PID2024-159504OB-I00, MICIU/AEI /10.13039/501100011033 and FEDER, UE.
\end{acknowledgments}
\section*{Declaration on Generative AI}
Artificial intelligence was used exclusively for spelling checks and grammar improvement in the preparation of this manuscript.





\bibliography{references}

\newpage

\appendix
\section{Full Ontology Requirements Specification Document (ORSD)}
\label{annex:orsd}

\begin{figure}[h]
    \centering
    \includegraphics[width=1\linewidth]{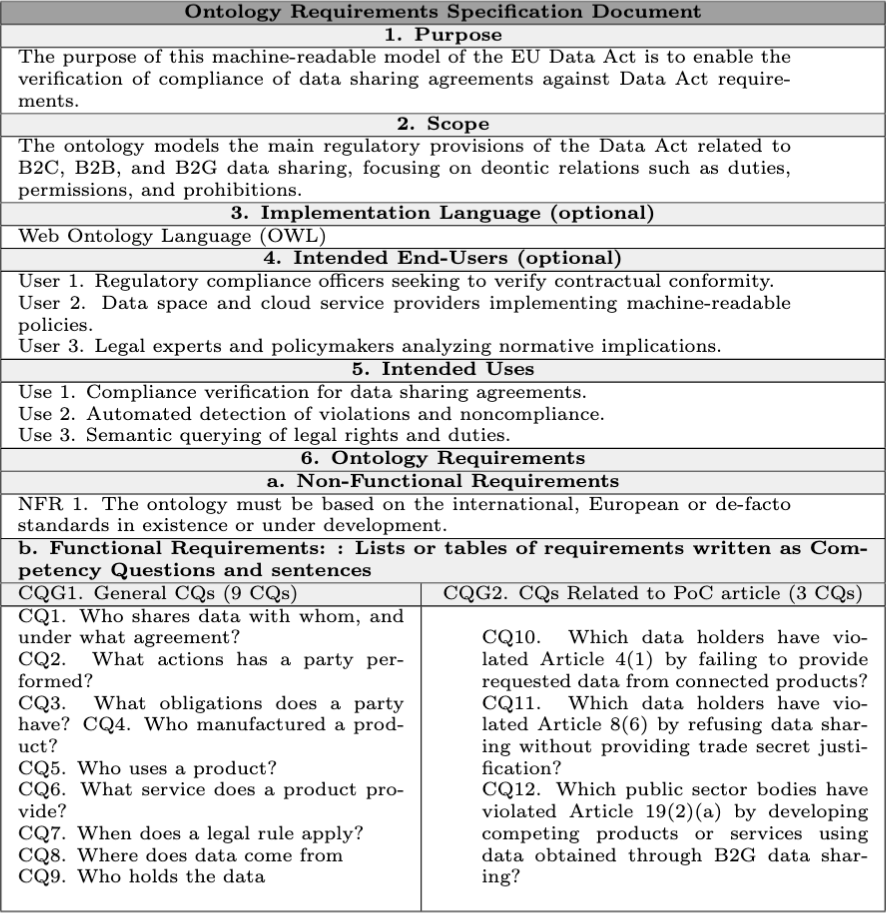}
    \caption{Complete Ontology Requirements Specification Document (ORSD) for DAOnt.}
    \label{fig:orsd}
\end{figure}

\end{document}